%% file: covert_ofdm.tex
\documentclass[conference,10pt]{IEEEtran}
\IEEEoverridecommandlockouts
\usepackage[utf8]{inputenc}
\usepackage[top=0.75in, bottom=1.1in, left=0.631in, right=0.631in]{geometry}
\usepackage[usenames,dvipsnames]{color}
\usepackage[disable]{todonotes}
\include{macros} 

\blindedfalse

\usepackage{hhline}

\usepackage{subcaption}

\usepackage[normalem]{ulem}

\usepackage{scalefnt}

\usepackage{glossaries}
\makeglossaries
\newacronym{NIC}{NIC}{network interface card}
\newacronym{OFDM}{OFDM}{orthogonal frequency-division multiplexing}
\newacronym{CP}{CP}{cyclic prefix}
\newacronym{CFO}{CFO}{carrier frequency offset}
\newacronym{DSSS}{DSSS}{direct-sequence spread spectrum}
\newacronym{SNR}{SNR}{signal-to-noise ratio}
\newacronym{SDR}{SDR}{software defined radio}
\newacronym{ICI}{ICI}{inter-carrier interference}
\newacronym{ISI}{ISI}{inter-symbol interference}
\newacronym{BPSK}{BPSK}{binary phase-shift keying}
\newacronym{QPSK}{QPSK}{quadrature phase-shift keying}
\newacronym{QAM}{QAM}{quadrature amplitude modulation}
\newacronym{LTF}{LTF}{long training field}
\newacronym{STF}{STF}{short training field}
\newacronym{AWGN}{AWGN}{additive white Gaussian noise}
\newacronym{BER}{BER}{bit error rate}
\newacronym{RSSI}{RSSI}{received signal strength indicator}
\newacronym{EVM}{EVM}{error vector magnitude}
\newacronym{LOS}{LOS}{line of sight}
\newacronym{FFT}{FFT}{Fast Fourier Transform}
\newacronym{IFFT}{IFFT}{Inverse Fast Fourier Transform}
\newacronym{SIG}{SIG}{signal field}
\newacronym{LNA}{LNA}{low-noise amplifier}
\newacronym{AGC}{AGC}{automatic gain control}
\newacronym{ADC}{ADC}{analog to digital converter}
\newacronym{B}{B}{residential}
\newacronym{D}{D}{typical office}
\newacronym{E}{E}{large office}
\newacronym{RMS}{RMS}{root mean square}
\newacronym{FCS}{FCS}{frame check sequence}
\newacronym{MAC}{MAC}{media access control}
\newacronym{MSDU}{MSDU}{MAC service data unit}
\newacronym{PSK}{PSK}{phase shift keying}
\newacronym{WARP}{WARP}{Wireless Open-Access Research Platform}
\newacronym{WEP}{WEP}{Wired Equivalent Privacy}
\newacronym{FCF}{FCF}{Frame Control Field}
\newacronym{SC}{SC}{subcarrier}
\newacronym{RTS}{RTS}{request to send}
\newacronym{CTS}{CTS}{clear to send}
\newacronym{FSK}{FSK}{frequency shift keying}
\newacronym{SIFS}{SIFS}{short inter-frame spacing}
\newacronym{FIR}{FIR}{finite impulse response}
\newacronym{IV}{IV}{initialization vector}

\usepackage[
  bookmarks=false,
  pdfpagelabels=false,
  hyperfootnotes=false,
  hyperindex=false,
  pageanchor=false,
  hidelinks,
  pdftex
]{hyperref}

\usepackage{amsmath}

\usepackage{tikz}
\usepackage{pgfplots}

\usepackage{dblfloatfix}

\newcommand{\pgffontsize}{\scriptsize}

\usepackage{nicefrac}

\usepackage{lipsum}

\usepackage{adjustbox}

\ifnum\pdfshellescape=1
\usepgfplotslibrary{external}
\else

\fi

\makeatletter
\def\ps@headings{%
\def\@oddhead{\mbox{}\scriptsize\rightmark \hfil \thepage}%
\def\@evenhead{\scriptsize\thepage \hfil \leftmark\mbox{}}%
\def\@oddfoot{}%
\def\@evenfoot{}}
\makeatother
\begin{document}
\begin{NoHyper}

\title{Practical Covert Channels for WiFi Systems}

\ifblinded
\numberofauthors{1}
\author{
	   \alignauthor Authors are blinded for review.\\
       \affaddr{\ }\\
       \affaddr{\ }\\
       \email{\ }\\
}
\else

\author{\IEEEauthorblockN{Jiska Classen {$^*$}\thanks{{$^*$} These authors contributed equally to this work.}, Matthias Schulz {$^*$}, and Matthias Hollick}
\IEEEauthorblockA{Secure Mobile Networking Lab\\ Technische Universität Darmstadt \\ \{jclassen, mschulz, mhollick\}@seemoo.tu-darmstadt.de\\ }
}
\fi

\maketitle

\begin{abstract}

Wireless covert channels promise to exfiltrate information with high bandwidth by circumventing traditional access control mechanisms.
Ideally, they are only accessible by the intended recipient and---for regular system users/operators---indistinguishable from normal operation.
While a number of theoretical and simulation studies exist in literature, the practical aspects of WiFi covert channels are not well understood.
Yet, it is particularly the practical design and implementation aspect of wireless systems that provides attackers with the latitude to establish covert channels: the ability to operate under adverse conditions and to tolerate a high amount of signal variations.
Moreover, covert physical receivers do not have to be addressed within wireless frames, but can simply eavesdrop on the transmission.
In this work, we analyze the possibilities to establish covert channels in WiFi systems with emphasis on exploiting physical layer characteristics.
We discuss design alternatives for selected covert channel approaches and study their feasibility in practice.
By means of an extensive performance analysis, we compare the covert channel bandwidth. 
We further evaluate the possibility of revealing the introduced covert channels based on different detection capabilities.

\end{abstract}

\input{introduction.tex}

\input{background.tex}

\input{overview.tex}

\input{implementation.tex}

\input{eval.tex}

\input{related.tex}

\input{conclusion.tex}

\section*{Acknowledgments}
\ifblinded
Acknowledgment has been blinded for review.\hfill\\\vspace*{4em}
\else
\scriptsize{
This work has been funded by the German Research Foundation (DFG) in 
the Collaborative Research Center (SFB) 1053 ``MAKI – Multi-Mechanism-Adaptation 
for the Future Internet'' and by LOEWE CASED. 
We thank Halis Altug, Athiona Xhoga and Stephan Pfistner for the
implementation of the first prototypes.}
\fi

\bibliographystyle{abbrv}
\bibliography{covert_ofdm}

\end{NoHyper}
\end{document}

%% file: macros.tex
\usepackage{ifthen}
\newboolean{draftversion}

\setboolean{draftversion}{false} 

\ifthenelse{\boolean{draftversion}}{\setlength{\paperwidth}{10.5in}\setlength{\marginparwidth}{1in}\setlength{\hoffset}{1in}}{}
\newcommand{\jiska}[1]{\ifthenelse{\boolean{draftversion}}{{\color{blue}{\textbar}\hspace{-.4em}\marginpar{\color{blue}{\emph{JC:}\\#1}}}}{}}
\newcommand{\ms}[1]{\ifthenelse{\boolean{draftversion}}{{\color{red}{\textbar}\hspace{-.4em}\marginpar{\color{red}{\emph{MS:}\\#1}}}}{}}
\newcommand{\mh}[1]{\ifthenelse{\boolean{draftversion}}{{\color{OliveGreen}{\textbar}\hspace{-.4em}\marginpar{\color{OliveGreen}{\emph{MH:}\\#1}}}}{}}

\newcommand*\phantomas[3][c]{%
\ifmmode 
\makebox[\widthof{$#2$}][#1]{$#3$}%
\else 
\makebox[\widthof{#2}][#1]{#3}%
\fi 
}

\newif\ifblinded
\blindedfalse

%% file: introduction.tex
\section{Introduction}

Wireless transmissions are broadly used, although properly securing them remains an issue.
Typically, applications resorting to communications are protected by
allowing information leakage only to authorized channels
such as data transmission to permitted applications.
Communication is often controlled by firewalls. However, potential adversaries might outsmart this protection and nevertheless leak information by setting up a covert channel; hidden within inconspicuous actions. For example, they could modify the application layer camouflaging text within an image on a shared storage, or they could alter the lower layers, e. g., within network protocols and timing.

When hiding information on upper layers only a few variations
such as using reserved bits or changing transmission timings are possible; since a firewall would easily any other type of modification \cite{zander2007}.
In contrast, physical wireless transmissions are not plain bits but symbols containing a high amount of noise and random signal variations.
Snatching raw data out of the air results in a very large amount of data compared to upper layer capturing,
still not revealing if the recording contained hidden information or not.
Regular WiFi receivers are designed to reconstruct the signal despite variations, hence their performance
does not significantly decrease when additional information is embedded.
Due to the wireless broadcast nature, frames can contain oblivious
sender and receiver addresses to not be suspicious to other network participants---and still be received by attackers.
For instance, an online banking application could establish a secure connection to a server but
maliciously publish login data over a covert wireless physical channel.

WiFi covert channels have been mostly studied in theory and simulation \cite{grabski2013}. 
Practical evaluations are scarce due to the complexity of modifying existing \glspl{NIC}, the work of Dutta et al. \cite{dutta2013} being an exception.
We close this gap: in our work, we evaluate practical covert channels on the \gls{WARP}\cite{warpProject} as well as off-the-shelf wireless \glspl{NIC} as legitimate receivers.
Using \gls{WARP}, we are able to utilize the same \gls{OFDM} modulation schemes as in 802.11a/g. 
Our covert channels can be easily adapted to \gls{OFDM}-based wireless communication systems such as LTE, DVB-T, and upcoming standards like LTE Advanced.
We aim at remaining compatible with the 802.11a/g standard and having little to no
performance decrease on off-the-shelf receivers.
Our contributions are as follows:
\begin{enumerate}
\item We analyze the IEEE 802.11a/g physical layer with respect to promising anchors for covert channels on frame level and symbol level.
\item We propose, analyze, and practically implement two novel covert channels. We study the performance in simulation and practice.
\item We analyze and improve two known covert channels; we practically implement them for the first time and study the performance in simulation and practice.
\item We compare the performance of all four covert channels and discuss practical limitations.
\end{enumerate}

This paper is structured as follows:
We introduce concepts behind WiFi covert channels in \autoref*{sec:background}.
System and security assumptions are defined in \autoref*{sec:overview}.
In \autoref*{sec:implementation}, covert channels and their performance in
practice are analyzed.
\autoref*{sec:evaluation} evaluates and discusses results.
In \autoref*{sec:relatedwork} we survey related work.
Finally, we conclude our results in \autoref*{sec:conclusion}.

%% file: background.tex
\section{Background}
\label{sec:background}

In the following, we introduce the concept of covert channels and
basic 802.11a/g physical layer operation.

\subsection{Covert Channels}

A first definition of covert channels is given in \cite{lampson1973} with a focus on information exchange between programs.
Channels are categorized as:

\begin{figure*}[!b]
\vspace{-1em}
\center
\includegraphics{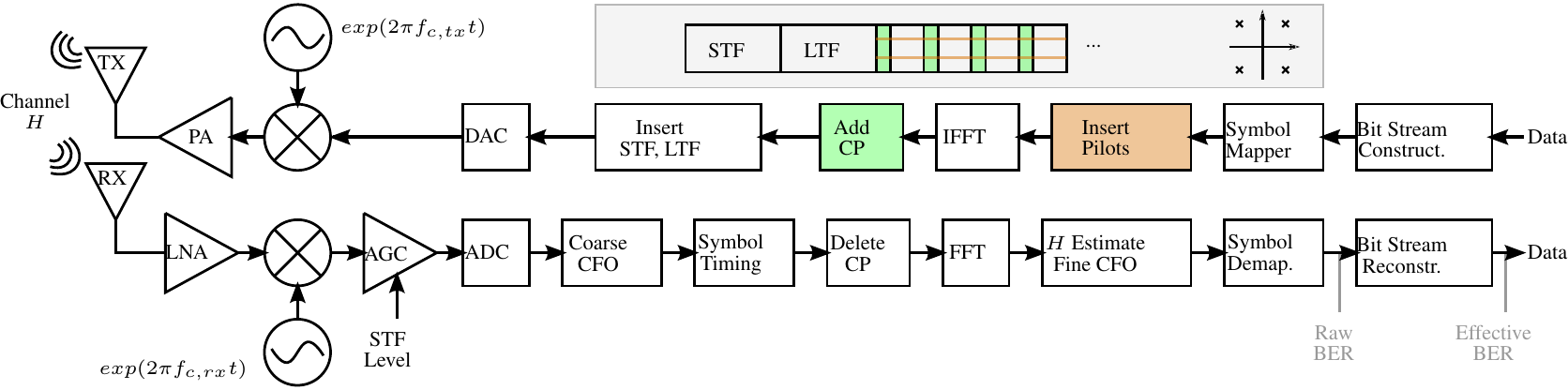}
\vspace{-15pt}
\caption{802.11a/g modulation and demodulation}
\label{fig:modem}
\end{figure*}

\begin{itemize}
\item \textit{legitimate}: information required to manage the program,
\item \textit{storage}: information provided to the program, however, attackers might have access to it, and
\item \textit{covert}: never intended for information exchange.
\end{itemize}
The idea of covert channels is similar to that of steganography, where messages are hidden within ordinary objects.
In case cryptography is forbidden within a network, covert channels can be used to hide
encrypted communication.

A covert channel consists of Alice, the sending attacker, who wants to communicate with Bob, the receiving attacker,
while being observed by Wendy, a warden.
Wendy's legitimate goal is to detect if Alice and Bob exchanged information.
In a wireless channel, positions of Alice, Bob, and Wendy are arbitrary---Wendy
might be closer to Alice than Bob.
Alice and Bob try to obscure the transmitted information to hinder Wendy from detection. 
Alice will typically send legitimate traffic to other stations and embed the covert channel. 
In contrast, communication between Alice and Bob is obvious in a cryptographic system and does not constitute an attack,
but Eve wants to illegitimately decipher their communication. 

Covert channels are implementable with and without keys.
Kerckhoff's law from cryptography is applicable to information hiding:
the system has to be secure when everything except the key is public.
Given this criteria, hiding information by relying on an unknown
embedding algorithm is insecure.
A wireless covert channel based on a public algorithm but private key
should be indistinguishable from noise.
Covert channels are often combined with cryptography to make information look like noise or to
add a further security measure.

\subsection{OFDM}

Physical layers of modern communication standards are based on \glsfirst{OFDM}.
To efficiently use the available
transmission bandwidth while still being able to correct channel distortions,
the transmission band is divided into \glspl{SC}. On each of these
subcarriers, symbols are transmitted by defining
amplitude and phase of the subcarrier frequencies for the duration of each
symbol $T_\text{sym}$. Limiting the length of each symbol leads to additional
frequency components in the form of sinc functions around each subcarrier. To
avoid \gls{ICI}, the spacing $\Delta f = 1/T_\text{sym}$ ensures that
each subcarrier is placed on the zero-crossings of the sinc functions of all
others, leading to orthogonality. During transmission,
the signal suffers from frequency-selective phase and amplitude changes
(fading), that can be corrected at the receiver. However, fading also implies a
delay spread leading to the reception of multiple time delayed copies of the
transmitted signal. To avoid \gls{ISI}, a guard interval is inserted between two
symbols, normally containing a continuation called \glsfirst{CP} of
the symbol.

\begin{figure}[!b]

    \vspace{-1em}
    \centering
	\includegraphics{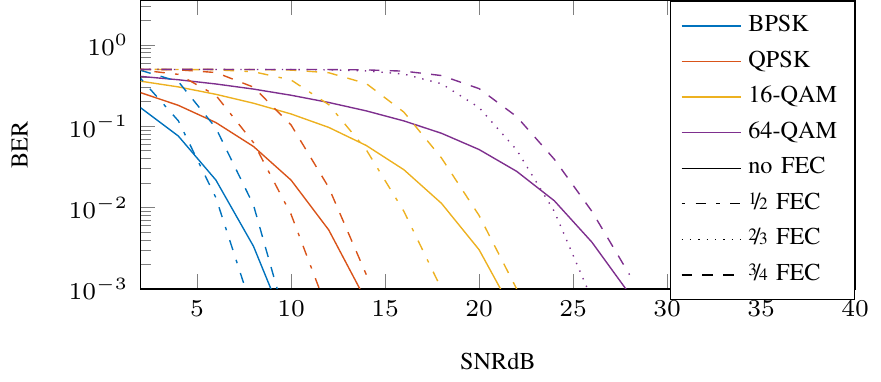}
    \vspace{-1em}
    \caption{BER baseline for WiFi frames before and after
    coding on an \gls{AWGN} channel with varying \gls{SNR}}
    \label{fig:ber-raw-effective-awgn}
    \vspace{-1em}
\end{figure}

\subsection{IEEE 802.11a/g physical layer}

In the following, we take a closer look at the frame structure
as well as at the \gls{OFDM}-based transceiver blocks of 802.11a/g systems as illustrated in \autoref*{fig:modem}.
The presented components are also required for more advanced standards.

For transmission, \gls{MAC} layer data bits are scrambled
to avoid consecutive ones or zeros, encoded for bit error correction, and
interleaved for distribution over multiple subcarriers. This bit stream is
mapped to symbols describing amplitude and phase of their
subcarrier. Depending on the modulation
order (bits per symbol) and the coding rate, eight gross transfer rates
between 6 and 54\,Mbps are defined in the WiFi standard \cite{ieee-80211}. In
\autoref*{fig:ber-raw-effective-awgn}, we illustrate the achievable \glspl{BER} on
a plain \gls{AWGN} channel before and after coding.
The used modulation scheme is documented in the \gls{SIG}
that is always encoded with 6\,Mbps.

Using the \gls{IFFT}, subcarriers are modulated according to symbol definitions
resulting in a time-domain signal in the baseband. Before upconversion to the
transmission frequency, a preamble
consisting of a \gls{STF} and a \gls{LTF} is prepended to every \gls{OFDM} frame. A
receiver needs the \gls{STF} to adjust the gain of its \gls{LNA}, and the
\gls{LTF} to estimate and correct channel effects on each subcarrier. 
Due to frequency differences in $f_\text{c,tx}$ and $f_\text{c,rx}$ as
well as frequency shifts due to the Doppler effect, \gls{CFO} occurs, which
breaks the orthogonality between subcarriers and hence requires correction.
Coarse \gls{CFO} correction makes use of the repetitive structure of either
\gls{STF} or \gls{LTF}, while fine \gls{CFO} correction uses pilot symbols that
are transmitted on four subcarriers of the \gls{OFDM} data symbols.

%% file: overview.tex
\section{System Overview}
\label{sec:overview}

In this section, we introduce a security model for wireless covert channels
and describe our measurement setup.

\subsection{Security Model}

To secure a system against covert channels, there are two main procedures:
either detecting or blocking them.
Blocking can be implemented by a wireless jammer~\cite{jamming,wifire}, though
jamming all wireless transmissions including legitimate ones is not an option. Since
there are no further processing steps between sending and receiving a signal,
there is no possibility to filter signal variations for covert channel blocking.
Detecting covert channels to take further actions such as jamming
or sender identification does not prevent legitimate wireless transmissions.
Sending attackers could be identified using localization methods or device
fingerprinting; however, fingerprints can be modified~\cite{rehman2014-lowend},
and localization requires multiple antennas.

A covert channel should be secure against detection, even if the information
hiding mechanism is known.
Detection security limits the capacity of covert channels.
Legitimate wireless transmissions containing a covert data have to be
indistinguishable from regular transmissions. Yet, the overall
wireless capacity is limited and a high capacity covert channel might noticeably
reduce legitimate throughput.

\textbf{Layer 1 detection.}
On the physical layer, detection requires \glspl{SDR}
or signal analyzers to capture the raw waveforms to measure \glspl{EVM},
\glspl{CFO}, and \glspl{SNR}.
A detector could compare these measurements to a benchmark set of typical values in wireless transmissions, and check which of them deviate significantly from a certain margin of statistical tolerance. Hence, an attacker should aim at keeping variations with respect to the signal relatively low, and let them only be remarkable in case a secret key is known, thus following Kerckhoff’s principle; which is reducing the actual possible covert channel throughput.

In this paper, we aim at showing the potential of practical and 802.11a/g compliant
covert channels. Providing an upper bound of performance, we do not implement statistical
detection countermeasures; however, we give some intuition into how they work in each channel covert description.

\textbf{Layer 2 detection.}
An upper layer detector is using off-the-shelf wireless \glspl{NIC}.
Even though this is not sufficient equipment to record ongoing transmissions on
the physical layer, information passed to upper layers might indicate whether a
covert channel is present.

Frames are validated on reception using the \gls{FCS} \cite{ieee-80211}. If
it fails, the frame is dropped by default. Higher layers can only rely on irregularities
in timing or throughput to detect covert channels.
In our evaluation, we enable the capture of those frames having failed \gls{FCS}  checks using radiotap headers~\cite{radiotap} to calculate actual \glspl{BER}.
Radiotap headers are supported by various chipsets and
forward additional information, such as
the transmission's center frequency, RF signal and noise power at the antenna, and the \gls{FCS}.
Detectors can correlate all this information, for instance, an increase of packet loss despite a constant RF signal power.
Still, radiotap headers do not provide the raw signal.
Applying Kerckhoff's law, information passed to upper layers is often insufficient for
detecting high throughput covert channels.

\begin{figure}[!t]
\centering
\begin{subfigure}[b]{0.43\linewidth}
\includegraphics[width=\linewidth]{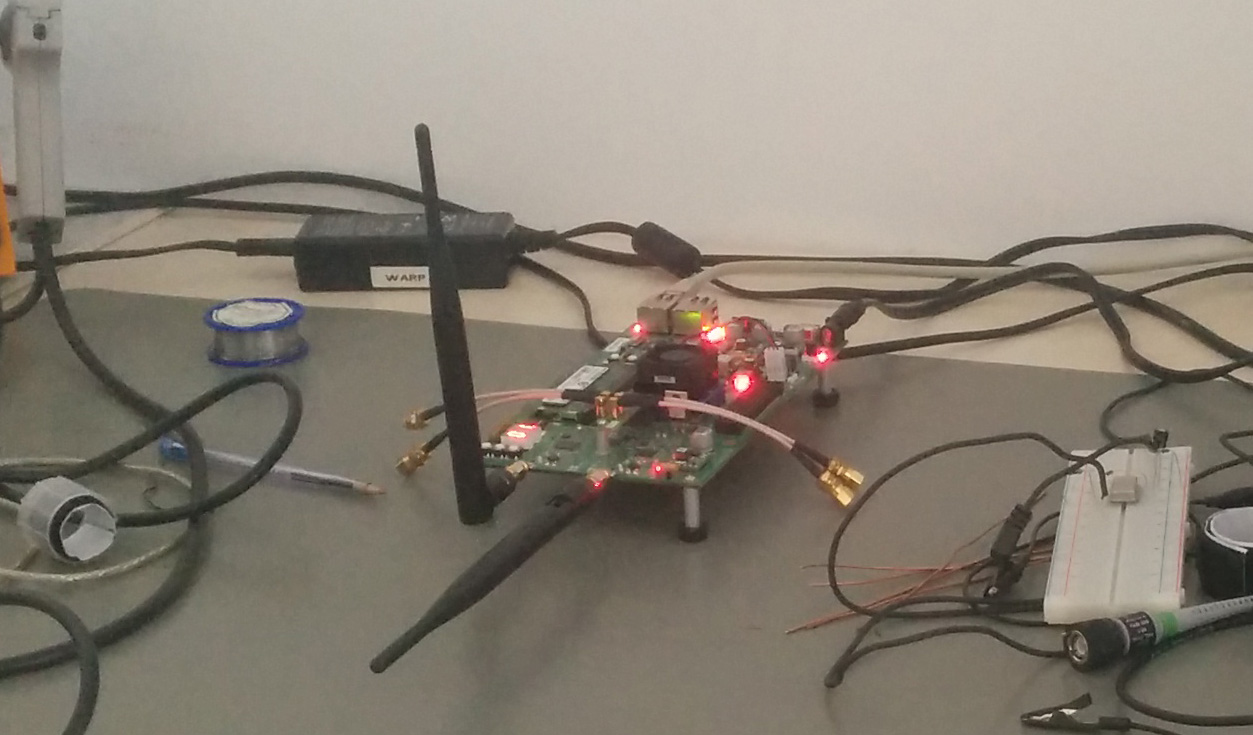}
\caption{Transmitter}
\vspace{-1.2em}
\label{fig:indoor_scenario}
\end{subfigure}\hspace{0.04\linewidth}\begin{subfigure}[b]{0.43\linewidth}
\includegraphics[width=\linewidth]{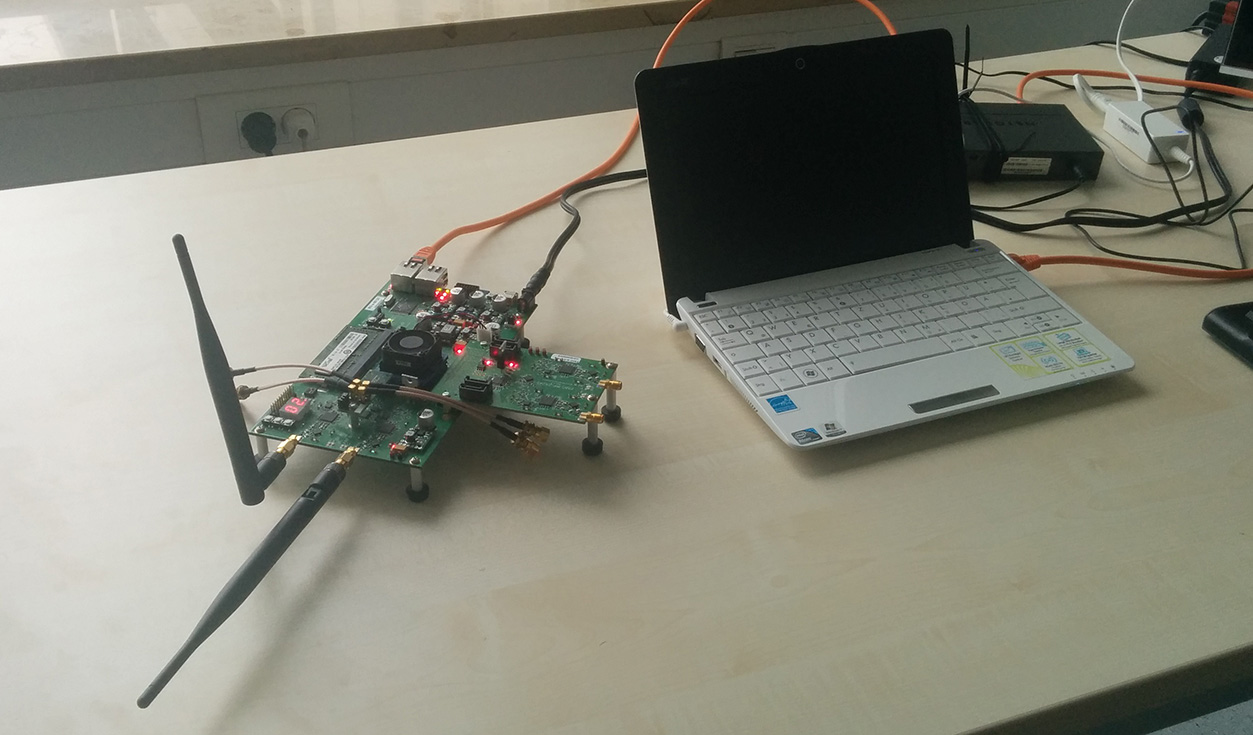}
\caption{Receivers}
\vspace{-1.2em}
\label{fig:indoor_scenario}
\end{subfigure}\vspace{1em}

\begin{subfigure}[b]{0.95\linewidth}
\centering
\begin{tikzpicture}[label distance=-2mm, font=\pgffontsize]
	\node[anchor=south west,inner sep=0] at (0,0)
	{\includegraphics[width=0.95\linewidth]{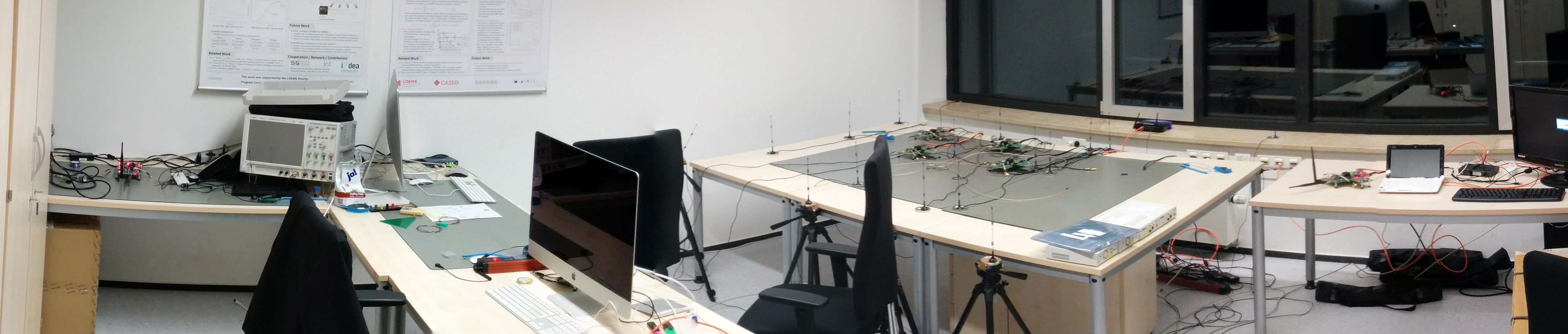}};
	
	\draw[black,thick,rounded corners] (0.41,0.69) rectangle (0.91,1.09);
	\draw[white,thick,rounded corners] (0.40,0.70) rectangle (0.90,1.10);
	\node[label={[black]0:\pgffontsize \textbf{WARP Transmitter}}] at (0.41,1.29)
	{}; 
	\node[label={[white]0:\pgffontsize \textbf{WARP Transmitter}}] at (0.4,1.3) {};
	
	\draw[black,thick,rounded corners] (6.51,0.69) rectangle (7.01,1.09);
	\draw[white,thick,rounded corners] (6.50,0.70) rectangle (7.00,1.10);
	\node[label={[black]0:\pgffontsize \textbf{WARP Receiver}}] at (5.01,1.29) {};
	\node[label={[white]0:\pgffontsize \textbf{WARP Receiver}}] at (5.0,1.3) {};
	
	\draw[black,thick,rounded corners] (7.01,0.69) rectangle (7.51,1.09);
	\draw[white,thick,rounded corners] (7.0,0.70) rectangle (7.50,1.10);
	\node[label={[black]0:\pgffontsize \textbf{Laptop Receiver}}] at (6.01,0.49)
	{}; \node[label={[white]0:\pgffontsize \textbf{Laptop Receiver}}] at (6.0,0.5)
	{};
	
	\draw[black,thick, <->] (1.00,0.90) -- (6.81,0.89);
	\draw[white,thick, <->] (1.00,0.90) -- (6.80,0.90);
	
	\node[label={[black]0:\pgffontsize \textbf{5\,m}}] at (3.51,1.05) {};
	\node[label={[white]0:\pgffontsize \textbf{5\,m}}] at (3.5,1.06) {};
\end{tikzpicture}
\caption{Panorama of the lab setup.}
\vspace{-1.2em}
\label{fig:indoor_scenario}
\end{subfigure}
\caption{Antenna setup for practical measurements.}
\vspace{-3em}
\label{fig:indoor_scenario}
\end{figure}

\subsection{Setup}
We analyze the performance of covert channels in simulation and practice using the following setups.

\subsubsection{Simulation}
We evaluate 
if the proposed covert 
channels are feasible utilizing diverse channel models: A (no fading), B
(residential), D (typical office), and E (large office) defined in
\cite{ieee-channels} and commonly used for WiFi simulations. To each model, we
add white Gaussian noise (\acrshort{AWGN}) and base our
results on 1000 Monte-Carlo simulations.

\subsubsection{Practical Setup}
Since simulations might disregard the behavior of real hardware, we evaluate all
covert channels in our lab environment (see \autoref*{fig:indoor_scenario}). This
evaluation is twofold.
We use \glspl{WARP} to transmit and receive 802.11g frames between Alice and Bob with
covert channels. On the receiver we can extract and analyze both, the
WiFi frame content and the covert channel. Hence, the \gls{WARP}
receiver can be considered as detector (Wendy) on Layer 1 as well as on the covert channel receiver (Bob). The signal processing
on both nodes is implemented in MATLAB which connects to the \glspl{WARP} using
WARPLab 7.5.0.

To analyze the effect on off-the-shelf WiFi devices, we use a laptop as detector (Wendy) on Layer 2 with a
Qualcom Atheros AR9285 Wireless Network Adapter (revision 01) that we run in monitor
mode with radiotap headers.

%% file: implementation.tex
\section{Covert Channels}
\label{sec:implementation}

In what follows, we present four practical covert channels for the physical layer of
802.11a/g. 
\begin{enumerate}
\item A covert channel utilizing the Short Training Field in combination with Phase Shift Keying (STF PSK).
\item A covert channel utilizing the Carrier Frequency Offset with Frequency Shift Keying (CFO FSK).
\item A covert channel using 802.11a/g with additional subcarriers conforming to the 802.11n spectrum mask (Camouflage Subcarriers).
\item A covert channel replacing parts of the \gls{OFDM} Cyclic Prefix (Cyclic Prefix Replacement).
\end{enumerate}
The schemes ``STF PSK'' and ``CFO FSK'' are new,
 ``Camouflage Subcarriers'' and ``Cyclic Prefix Replacement'' are extensions and
improvements to \cite{hijaz2010} and \cite{grabski2013}, respectively.
To the best of our knowledge, none of the approaches were put into practice before.

\subsection{Short Training Field with Phase Shift Keying (STF PSK)}
\label{sec:stf_psk}

Each 802.11a/g frame starts with the same \gls{STF} in the preamble, which is
used for frame detection, \gls{AGC}, and coarse \gls{CFO}
estimation. \gls{STF} manipulations must preserve these capabilities at
the receiver, otherwise the signal can not be demodulated.
Implementing a covert channel in the \gls{STF} allows to insert one symbol per
WiFi frame that is impossible to block even after detection.

\subsubsection{Implementation}
\begin{figure}[!t]
\center

\begin{subfigure}[b]{0.48\linewidth}
\centering
\includegraphics{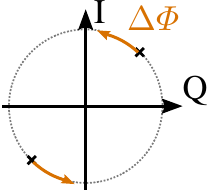}
\caption{theoretical}
\label{fig:stf-psk}
\end{subfigure}\hspace{0.04\linewidth}%
\begin{subfigure}[b]{0.48\linewidth}
\centering
\includegraphics{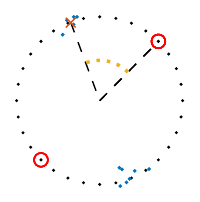}
\caption{measurement}
\label{fig:stf-psk-mes}
\end{subfigure}
\vspace{-2em}
\caption{STF PSK symbols are shifted by $\Delta \phi$ to encode
bits.}
\vspace{-1.2em}
\end{figure}

The \gls{STF} contains \gls{BPSK} symbols that are shifted by 45$^\circ$ as
illustrated in \autoref*{fig:stf-psk}. We insert our covert channel by
introducing an additional phase shift $\Delta\phi$ into all \gls{STF} symbols.
As phase shifts do not change the power and correlation properties of the
\gls{STF}, it can still be used for \gls{AGC} and packet detection.
Additionally, the periodicity required for \gls{CFO} correction is preserved.

Per WiFi frame, we insert one phase shift. Depending on the number of bits we
intend to encode, we vary the number of possible phase shift values mapped to bits using Gray coding.
To extract the covert channel information, the receiver needs to compensate the
channel effects in the \gls{STF} using the \gls{LTF} channel estimation.

Then, $\Delta\phi$ can be extracted and demapped to bits. In
\autoref*{fig:stf-psk-mes}, we illustrate this process with 32 possible phase
shifts (32-PSK illustrated by black dots). The red circles mark the original
\gls{STF} symbol positions, and the cloud of blue dots are the received \gls{STF}
symbols from which we extract the phase difference to the original symbol
positions.

\subsubsection{Performance}

\begin{figure}[!t]
    \centering
	\includegraphics{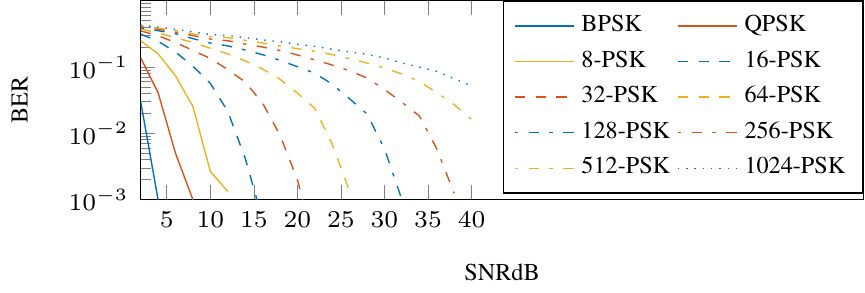}    
    \caption{Raw \gls{BER} of the covert channel implemented as \gls{STF}
    \gls{PSK} scheme over an \gls{AWGN} channel for different amounts of bits
    per frame.}
    \vspace{-1.2em}
    \label{fig:ber-raw-covert-stf-psk}
\end{figure}

Assuming we transmit WiFi frames with \gls{STF} \gls{PSK} covert
data over \gls{AWGN} channels without fading, we can reach the \glspl{BER}
illustrated in \autoref*{fig:ber-raw-covert-stf-psk}. The more bits we encode in
the \gls{STF}, the smaller the distance between the phase steps. This results in
an increased \gls{BER}. For a typical 25\,dB \gls{SNR}, 6 covert bits per
\gls{STF} can be hidden with less than 0.1\% covert channel \gls{BER}.

\begin{figure}[!t]
\begin{subfigure}[b]{\linewidth}
    \centering
    \includegraphics{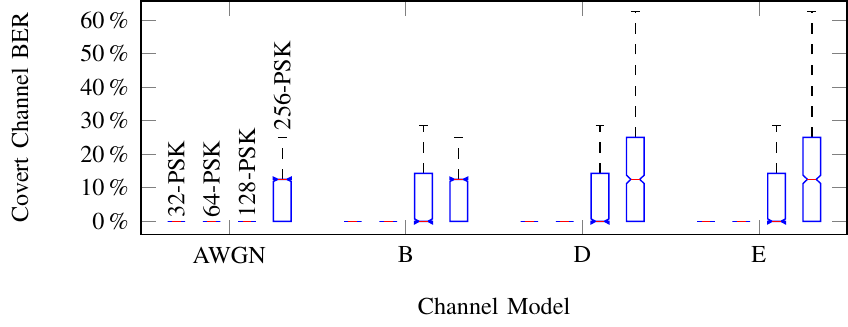}
	\caption{Simulation with 24\,Mbps WiFi frames (SNR\,=\,25\,dB).}
    \vspace{-1.2em}
    \label{fig:ber-cc-sim-stf-psk}
\end{subfigure}\vspace{1em}
\begin{subfigure}[b]{\linewidth}
    \centering
	\includegraphics{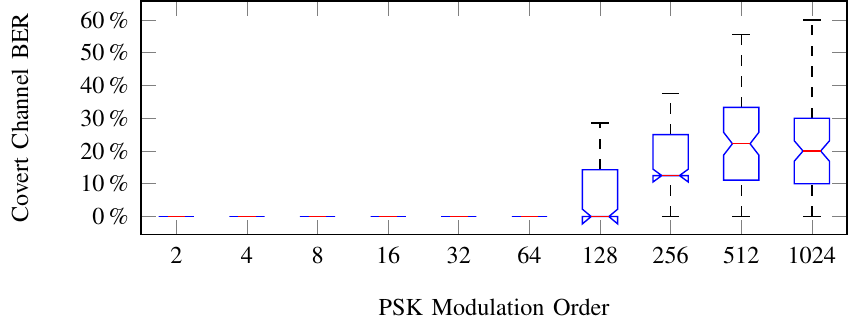}
    \caption{WARP-to-WARP measurement with 54\,Mbps WiFi frames.}
    \vspace{-1.2em}
    \label{fig:ber-cc-mes-stf-psk}
\end{subfigure}
\caption{BER of the STF PSK covert channel.}
\vspace{-1.2em}
\label{fig:ber-cc-stf-psk}
\end{figure}

To evaluate the \gls{STF} \gls{PSK} performance in fading channels, we
perform simulations with channel models B, D, and E with a fixed \gls{SNR} of
25\,dB introduced by \gls{AWGN}. In \autoref*{fig:ber-cc-sim-stf-psk}, we
illustrate the results from 32-PSK (5 bits/symbol) to 256-PSK (8 bits/symbol).
We observe that transmissions up to 64-PSK modulation are always error free, while
higher modulation orders result in more bit errors, especially when the effects
of fading increase.
In our lab, we measure that all modulation orders up to 128-PSK
have low median error rates as illustrated in \autoref*{fig:ber-cc-mes-stf-psk}.
We conclude that one can transfer roughly 6 to 7 bits per frame with very low \gls{BER}.

The achievable throughput of the covert channel strongly depends on the number
of WiFi frames transmitted per second.
For this scheme, short frames such as ACK and CTS (both 14 bytes long and sent at least at 36\,Mbit/s) are ideal, since one 4\,$\mu$s  long OFDM-symbol sequence holds the complete \gls{MAC} layer payload. Note that increasing frame rates without a plausible reason might help Wendy to detect information exchanges.
Combined with \gls{STF} (4\,$\mu$s), \gls{LTF} (8\,$\mu$s) and signal field
(4\,$\mu$s), the complete frame is 16\,$\mu$s long; resulting in a gross frame
rate of 62,500 frames/s. Using 64-PSK the \gls{STF} \gls{PSK} covert channel
achieves a gross bitrate of 375\,kbit/s.

\begin{figure*}[!b]
\vspace{-1em}
\begin{minipage}[b]{.32\textwidth}
	\includegraphics{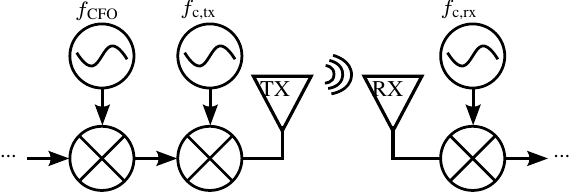}
	\caption{We introduce artificial \gls{CFO} $f_\text{CFO}$ into each
	\gls{OFDM} symbol in the baseband.}
	\label{fig:cfo}
\end{minipage}\hspace{0.04\linewidth}\begin{minipage}[b]{.30\textwidth}
	\includegraphics{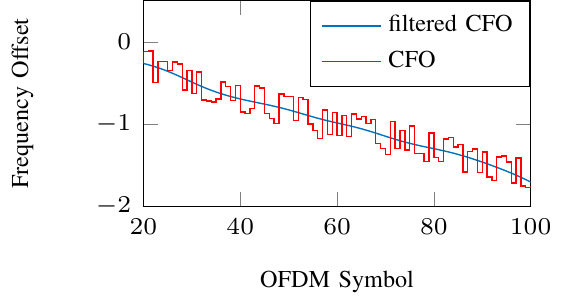}
    \vspace{-1.5em}
    \caption{Frequency offset measurement of each received \gls{OFDM} symbol
    showing the binary shift keying modulation.}
    \label{fig:frequency-offset-illustration}
\end{minipage}\hspace{0.04\linewidth}\begin{minipage}[b]{.30\textwidth}
	\includegraphics{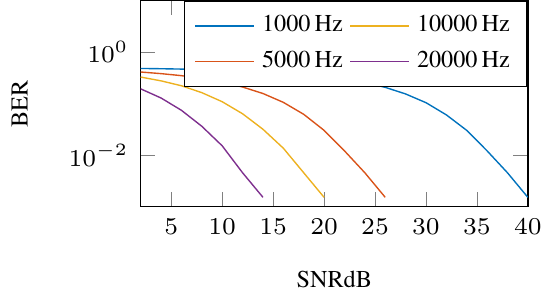}
    \vspace{-1.5em}
    \caption{Raw \gls{BER} of the \gls{CFO} \acrshort{FSK} covert channel over
    \gls{AWGN} channels with different $\Delta_\text{CFO}$.}
    \label{fig:ber-covert-raw-cfo-fsk}
\end{minipage}
\end{figure*}

\subsubsection{Detection}

\textbf{Layer 1.} A physical layer detector needs to perform the same steps as
the covert channel receiver mentioned above. Those steps are not accomplished in
regular WiFi receivers and require a custom \gls{SDR}-based implementation or a
spectrum/signal analyzer. To lower the detection probability, a transmitter can
map bits only to small phase changes, which results in reduced covert channel
throughput. As the secret information is already transmitted before it can be
detected, a wireless jammer cannot be used to block the covert channel
transmission without destroying every WiFi frame.

\textbf{Layer 2.} As mentioned above, a phase shift in the \gls{STF} does not
influence the functionality of the \gls{STF} at the receiver. To verify this, we
compared \glspl{BER} of received WiFi frames with and without covert channel and
were not able to distinguish between them. Neither in simulation, nor in practice 
when receiving with a \gls{WARP} or off-the-shelf WiFi card.

\subsection{Carrier Frequency Offset with Frequency Shift Keying (CFO FSK)}
\label{sec:cfo_fsk}

A WiFi baseband signal is upconverted to the carrier
frequency $f_c$ with $f_\text{c,tx}
\approx f_c$ and downconverted using $f_\text{c,rx} \approx
f_c$ (see \autoref*{fig:cfo}). Their difference results in \gls{CFO}, which needs to be corrected, together with
additional \gls{CFO} due to the Doppler effect.
WiFi receivers are capable of correcting \glspl{CFO} by
tracking the pilots that are inserted into each \gls{OFDM} symbol. We introduce an artificial
$f_\text{CFO}$ at the transmitter as covert channel. Regular WiFi receivers silently correct
$f_\text{CFO}$, while covert channel receivers can extract the hidden
information.

\subsubsection{Implementation}
To encode bits, the transmitter maps them
to the two frequencies $\pm\Delta_\text{CFO}$, each with a symbol length of an \gls{OFDM} symbol
(4\,$\mu$s). The resulting complex waveform is multiplied with the
time-domain signal of the WiFi frames in the baseband. This shifts each
\gls{OFDM} symbol by $\pm\Delta_\text{CFO}$ in the
frequency-domain, depending on the encoded bit.

A covert channel receiver estimates the phase shifts of the pilot symbols for
each \gls{OFDM} symbol, as illustrated in
\autoref*{fig:frequency-offset-illustration}. The covert \gls{CFO} changes are
superimposed by an additional slowly varying \gls{CFO}. To extract
bits despite further \gls{CFO} components, the receiver first lowpass filters the \gls{CFO} estimate and uses it as a
threshold for a hard decision decoder. The six outer
bits on both sides are discarded as they contain many bit errors. The lowpass
filter is implemented as 20-tap \gls{FIR} filter, which requires at least 60
\gls{OFDM} symbols to work correctly.

\subsubsection{Performance}
\begin{figure}[!t]
    \centering
	\includegraphics{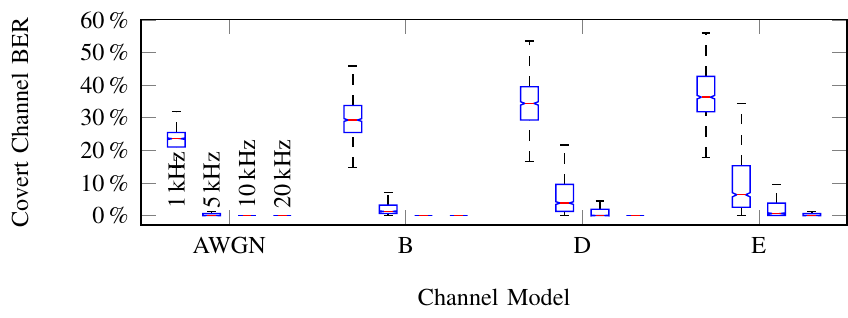}    
    \caption{CFO FSK covert channel simulation with 24\,Mbps WiFi frames (SNR\,=\,25\,dB).}
    \vspace{-1.5em}
    \label{fig:ber-cc-sim-cfo-fsk}
\end{figure}

In the simulations we add a fixed 50\,kHz \gls{CFO} for both \gls{AWGN}
and fading channels as well as a 15\,Hz maximum Doppler spread for the fading
channels B, D, and E, representing environmental movement.
The resulting \gls{AWGN} covert rates for different
$\Delta_\text{CFO}$ values in
\autoref*{fig:ber-covert-raw-cfo-fsk} show that stronger \gls{CFO} changes enhance the covert channel.
As illustrated in \autoref*{fig:ber-cc-sim-cfo-fsk}, stronger multipath effects lead
to higher covert \glspl{BER}. Especially in the model E,
a $\Delta_\text{CFO}$ of more than 10\,kHz is required to keep the
\glspl{BER} low.
In WARP-to-WARP measurements with 54\,Mbps WiFi frames, for $\Delta_\text{CFO}$=1\,kHz, the average covert BER is 15\%---for
$\Delta_\text{CFO}\geq$5\,kHz no errors occur, which is comparable to the \gls{AWGN} simulation results.

\begin{figure}[!t]
\begin{subfigure}[b]{\linewidth}
    \centering
	\includegraphics{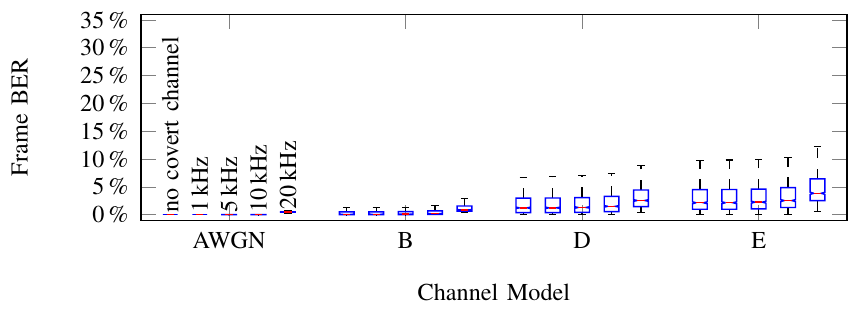}
    \caption{Simulation with 24\,Mbps WiFi frames (SNR\,=\,25\,dB).}
    \vspace{-1.5em}
    \label{fig:ber-wifi-sim-cfo-fsk}
\end{subfigure}\vspace{1em}
\begin{subfigure}[b]{\linewidth}
    \centering
    \includegraphics{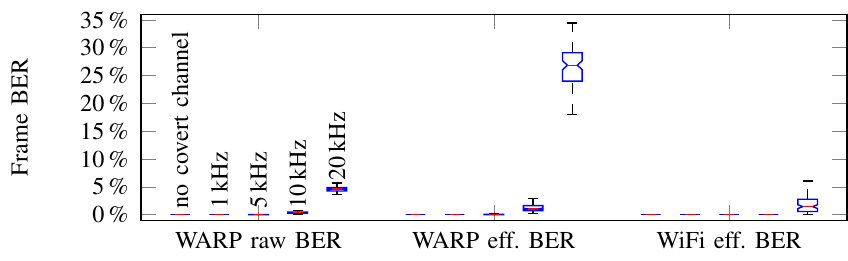}
    \caption{WARP-to-WARP/Laptop 54\,Mbps legitimate receiver.}
    \vspace{-1.5em}
    \label{fig:ber-wifi-mes-cfo-fsk}
\end{subfigure}
\caption{CFO FSK BER at the legitimate receiver.}
\vspace{-2.5em}
\label{fig:ber-wifi-cfo-fsk}
\end{figure}

The \glspl{BER} of the WiFi frames in both simulation
(\autoref*{fig:ber-wifi-sim-cfo-fsk}) and practice
(\autoref*{fig:ber-wifi-mes-cfo-fsk}) show that---up to 10\,kHz
$\Delta_\text{CFO}$---there is almost no increase in the \glspl{BER} at the detector.
To avoid detection, the lowest working
$\Delta_\text{CFO}$ should be chosen, which is 5\,kHz in our lab.
By encoding 1\,bit per 4\,$\mu$s \gls{OFDM} symbol,
the covert throughput is 250\,kbit/s.

\subsubsection{Detection}

\textbf{Layer 1.} Every WiFi receiver estimates and corrects \glspl{CFO}.
However, those measurements are normally directly discarded during signal
processing. As shown in \autoref*{fig:frequency-offset-illustration}, receivers
capable of analyzing \gls{CFO} changes over time can directly detect the binary
pattern. Using lower $\Delta_\text{CFO}$ values hardens detection but 
increases error probabilities on the covert channel. 

\textbf{Layer 2.} Our simulated and practical results in
\autoref*{fig:ber-wifi-cfo-fsk} show that large \gls{CFO} changes drastically
increase \glspl{BER} in all channel models.
However, in our setup 5\,kHz $\Delta_\text{CFO}$ is sufficient for covert
transmissions without increasing errors in the WiFi frame reception.
Furthermore, $\Delta_\text{CFO}$ could slowly be increased
to stealthily reach a working point to prevent detectable sudden \gls{BER} changes.
Hence, \gls{CFO} \gls{FSK} can be undetectable on Layer 2, if configured carefully.

\subsection{Camouflage Subcarriers}
\label{sec:cs}

\begin{figure}[!t]
    \centering
	\includegraphics{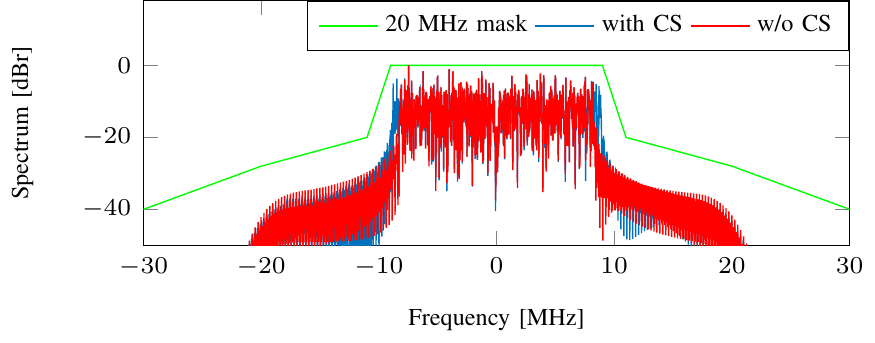}
    \caption{Spectra of both regular 802.11g frames and camouflage subcarrier frames fit into 20\,MHz WiFi channels.}
    \vspace{-1.2em}
    \label{fig:hips-spectra}
\end{figure}

The camouflage subcarrier covert channel hides information in subcarriers used
in other protocol variants. In 802.11a/g, 52 subcarriers are used for 48 data and
4 pilot transmissions, while 802.11n utilizes 56 subcarriers in the same band. The additional 4 subcarriers can be utilized in 802.11a/g
transmissions as covert channel. At plain sight
the spectra look like valid 802.11n frames (as depicted in
\autoref*{fig:hips-spectra}). A regular 802.11a/g/n WiFi receiver 
does not sense the number of used subcarriers, but only checks the
signal field at the beginning frame and continues decoding
according to the 802.11a/g standard, simply ignoring camouflage subcarriers.
Using additional subcarriers was proposed in \cite{hijaz2010}, yet, without the constraint to
mimic another protocol version.

\subsubsection{Implementation}
We replace the 802.11a/g \gls{LTF} with the 802.11n HT-LTF, which is still
correlating with the LTF, thus allowing a proper timing synchronization at
the receiver. Additionally, the covert receiver can estimate the
channel effects of the camouflage subcarriers.

\subsubsection{Performance}
\begin{figure}[!t]
\vspace{-.1em}
    \centering
	\includegraphics{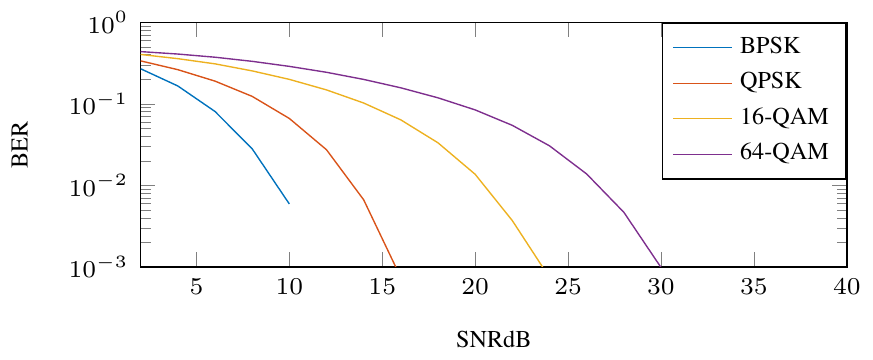}
    \caption{Raw \gls{BER} of the camouflage subcarriers
    over AWGN.}
    \vspace{-1.5em}
    \label{fig:ber-covert-raw-hips}
\end{figure}

\begin{figure}[!t]
    \centering
	\includegraphics{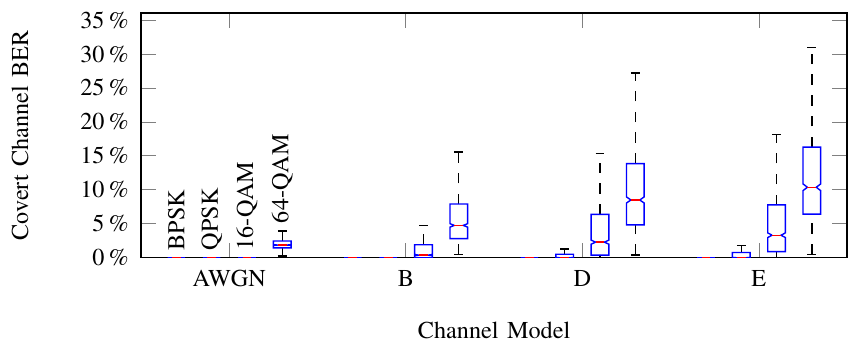}
	\caption{BER of covert camouflage subcarriers in simulation
with 24\,Mbps WiFi frames (SNR\,=\,25\,dB).}
	\vspace{-1em}
	\label{fig:ber-cc-sim-sc}
\end{figure}

When comparing \autoref*{fig:ber-covert-raw-hips} to \autoref*{fig:ber-raw-effective-awgn},
it is obvious that the covert subcarriers perform very similar to the normal subcarriers.
Depending on channel effects and output filters, it might happen that the outer subcarriers
have a slightly different performance, though. 
Covert subcarrier performance for different channel models is depicted in \autoref*{fig:ber-cc-sim-sc}.
Assuming camouflage and normal subcarrier performance are similar,
the covert channel performance is 8.3\,\% of the normal
channel throughput.

In our experiments, we vary the rate of the
camouflage subcarriers, while keeping the rate of the regular WiFi data fixed.
\autoref*{fig:ber-cc-sim-sc} compares simulation results of camouflage subcarriers.
Experimental results are not illustrated---the WARP-to-WARP channel in our lab
is quite good and no errors occur in the camouflage subcarriers for all modulation orders.

\subsubsection{Detection}

\textbf{Layer 1.} A Layer 1 detector that can decode the signal field is able
to determine if the number of subcarriers within the signal is correct.
However, only checking the spectrum will not reveal the covert channel, as it is
still valid and conforms to the standard 802.11n.

\textbf{Layer 2.} A Layer 2 detector has insufficient information
since neither normal subcarrier performance decreases nor
interference with neighboring channels occurs.
Even further subcarriers can be used to increase covert channel throughput as
long as the neighboring channels do not overlap, but this could be easily detected
on Layer 1. Our results show that adding camouflage subcarriers does neither
increase \glspl{BER} in simulation nor in practice.

\subsection{Cyclic Prefix Replacement}
\label{sec:cprep}
\label{sec:cp}
Multipath effects and timing offsets during demodulation cause overlapping
\gls{OFDM} symbol parts, called \acrfull{ISI}.
In 802.11a/g, a \gls{CP} is prepended to symbols in order to reduce \gls{ISI}.
At reception, this \gls{CP} is not decoded.
Nevertheless, the \gls{CP} might still be larger than the actual \gls{ISI} and, hence, can be used
as a covert channel.

\begin{figure}[!t]
\centering

\begin{subfigure}[b]{0.27\textwidth}
\flushleft
\includegraphics{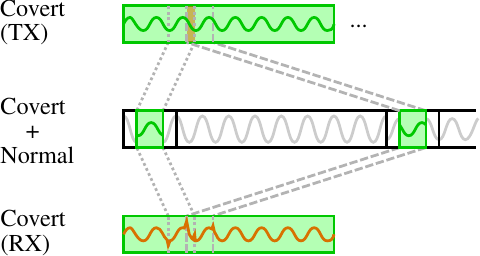}
\vspace{-10pt}
\caption{\acrshort{CP} cutting and embedding within one subcarrier in the time-domain.}
\label{fig:cutting}
\end{subfigure}\hspace{10pt}\begin{subfigure}[b]{0.18\textwidth}

\flushright
\includegraphics{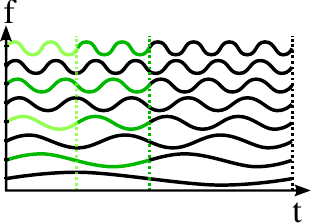}
\caption{Effect of a smaller \acrshort{FFT} size for covert symbols within one \acrshort{CP}.}
\label{fig:short-fft}
\end{subfigure}
\vspace{-1em}
\caption{\acrshort{CP} replacement methods compared.}
\vspace{-5em}
\end{figure}

\begin{figure}[!b]
    \centering
	\includegraphics{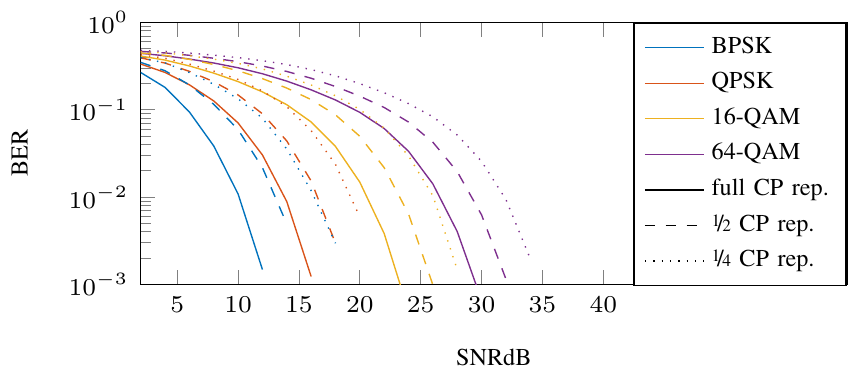}    
    \caption{Raw \gls{BER} of the Cyclic Prefix Replacement covert channel
    over AWGN channels.}
    \label{fig:ber-raw-stf-psk}
\end{figure}

A simulation in~\cite{grabski2013} replaces the complete \gls{CP} with covert
symbols. This results in a normal channel with up to 54\,Mbit/s according
to 802.11a/g and an additional covert channel achieving 13.5\,Mbit/s, since
the \gls{CP} length is $\nicefrac{1}{4}$ of the normal symbol. The channel
performs well as the simulations are limited to \gls{AWGN} channels with neither fading nor \gls{ISI}. Hence, the \gls{CP} is not required at all. In
practice, we could not reproduce such optimistic results.

\subsubsection{Implementation}

There are basically two ways of embedding data in the \gls{CP}.
In the first approach, four \glspl{CP} are combined to obtain a symbol of regular
length. In a practical channel instead of the \gls{AWGN} channel
proposed in ~\cite{grabski2013}, fading effects disturb samples near to
concatenation points. A solution is shown in \autoref*{fig:cutting}, where the
covert symbols are distributed to multiple \glspl{CP} with some overlapping
samples. First simulation results, however, show that more concatenations lead
to more disturbances (e.g. due to the Doppler effect) making this approach
impractical.

The second approach decreases the \gls{FFT} size to a maximum of 
the actual CP length, automatically leading to less subcarriers as depicted in 
\autoref*{fig:short-fft}.
Even though only $\nicefrac{1}{4}$ of the subcarriers are used in a 16-point
\gls{FFT} compared to the normal symbol's 64-point \gls{FFT}, 12 symbols are
usable by replacing the full \gls{CP}. Using four \glspl{CP}, 48 symbols can
be used for data transmission---analogous to the first approach. 
To reduce the \gls{ISI} with regular \gls{OFDM} symbols, the covert channel
\gls{FFT} size can be reduced to 8, 4, or 2 at the cost of covert throughput.
Prepending a \gls{CP} to the covert channel in the \gls{CP}
(called CPCP) even removes \gls{ISI} inside the covert channel. In our experiments,
we add a CPCP of 2 samples to the $\nicefrac{1}{2}$ CP replacement scheme.

\subsubsection{Performance}
\begin{figure}[!t]
\vspace{-.2em}
\begin{subfigure}[b]{\linewidth}
    \centering
	\includegraphics{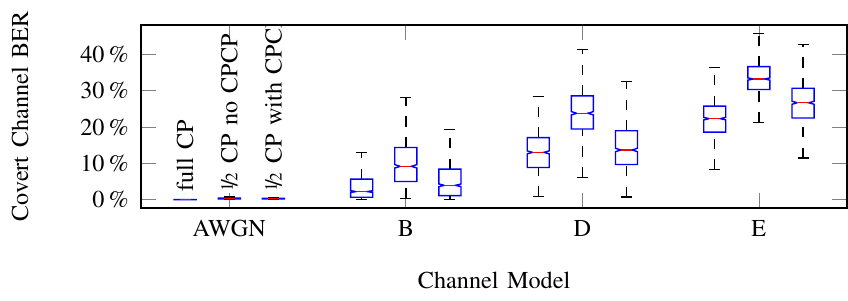}
    \vspace{-.5em}
    \caption{Simulation with 24\,Mbps WiFi frames (SNR\,=\,25\,dB).}
    \label{fig:ber-cc-sim-cpreplacement}
\end{subfigure}\vspace{.7em}
\begin{subfigure}[b]{\linewidth}
    \centering
	\includegraphics{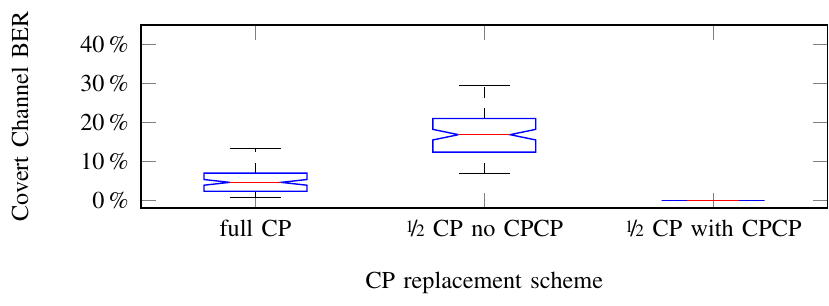}
    \vspace{-.5em}
    \caption{WARP-to-WARP measurement with 54\,Mbps WiFi frames.}
    \label{fig:ber-cc-mes-cpreplacement}
\end{subfigure}
\vspace{-2em}
\caption{BER of covert \gls{CP} replacement.}
\label{fig:ber-cc-cpreplacement}
\vspace{-1em}
\end{figure}

The performance of the \gls{CP} replacement covert channel is very high.
\autoref*{fig:ber-raw-stf-psk} compares \glspl{BER} for different \gls{CP} replacement strategies in an \gls{AWGN} channel.
Replacing shorter parts of the \gls{CP}
results in more errors. Adding a CPCP does not help in an \gls{AWGN} channel because the channel does
not introduce \gls{ISI}.
In contrast, in the multipath channel simulations illustrated in \autoref*{fig:ber-cc-sim-cpreplacement}, the CPCP significantly decreases the covert channel \glspl{BER}.
In our lab environment, the CPCP is required and very effective: it reduces the \gls{BER} to 0\% as shown in \autoref*{fig:ber-cc-mes-cpreplacement}.
Depending on the actual amount of multipath effects, a higher CPCP length is reasonable.

Throughput of full CP replacement is 25\,\% of the corresponding WiFi
frame throughput, if multipath effects are neglected. For $\nicefrac{1}{2}$ CP
replacement, the maximum throughput is reduced to 12.5\% of the WiFi frame
throughput. Hence, even with the CPCP improvement for less transmission errors,
this covert channel has good performance.

\begin{figure}[!t]
\vspace{-.2em}
\begin{subfigure}[b]{\linewidth}
    \centering
	\includegraphics{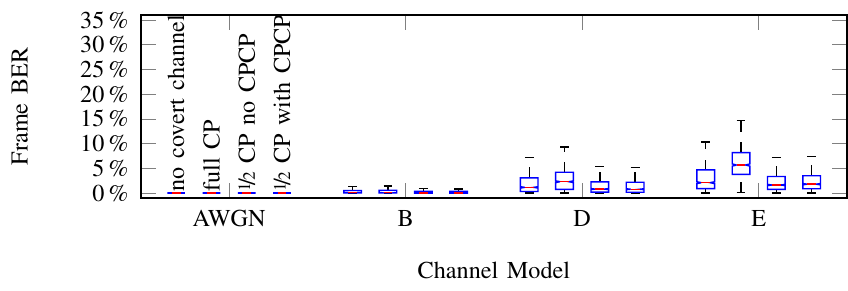}
    \vspace{-.5em}
    \caption{Simulation with 24\,Mbps WiFi frames (SNR\,=\,25\,dB).}
    \label{fig:ber-wifi-sim-cpreplament}
\end{subfigure}\vspace{.7em}
\begin{subfigure}[b]{\linewidth}
    \centering
	\includegraphics{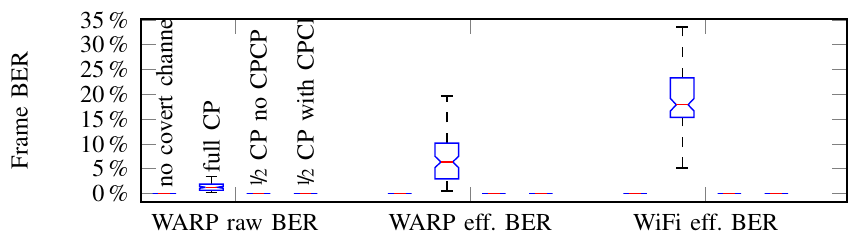}
    \vspace{-.5em}
    \caption{WARP-to-WARP/Laptop measurement with 54\,Mbps WiFi frames.}
    \label{fig:ber-wifi-mes-cpreplament}
\end{subfigure}
\vspace{-2em}
\caption{BER of WiFi frames with \gls{CP} replacement at legitimate receivers.}
\label{fig:ber-wifi-cpreplament}
\end{figure}

\begin{figure}[!t]
\vspace{-1em}
    \centering
	\includegraphics{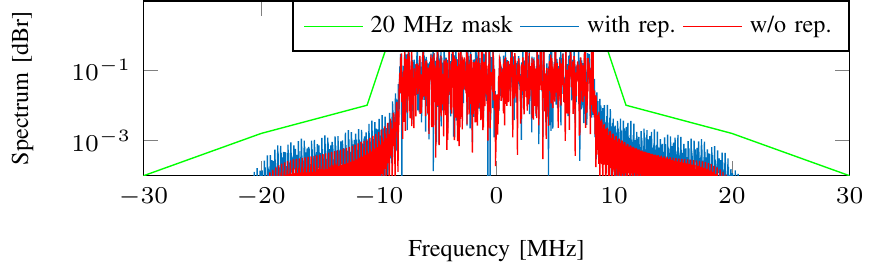}
    \vspace{-1.5em}
    \caption{The spectrum of CP Replacement frames has 
    higher out-of-band transmissions than regular frames.}
    \label{fig:cp-spectra}
\vspace{-2em}
\end{figure}

\begin{table*}[!b]
\vspace{-1em}
\centering
\captionsetup{justification=centering}

\begin{tabular}{p{0.10\textwidth} c p{0.68\textwidth} } \hline  
    Covert Channel & Section & Conclusion \\ \hline

    STF PSK & Sec. \ref{sec:stf_psk}
    & Introduces phase shift to \gls{STF}; immune to reactive jamming; no
    influence on WiFi \gls{BER}; 1 PSK symbol per frame; max. covert rate 375\,kBit/s for 64-PSK
    \\[.1cm]
        
    CFO FSK & Sec. \ref{sec:cfo_fsk} 
    & Introduces artificial \gls{CFO}; tunable for no influence on Wifi
    \gls{BER}; 1 bit per OFDM symbol; max. covert rate 250\,kBit/s for 5\,kHz FSK
    \\[.1cm]
    
	Camouflage\newline Subcarriers  & Sec. \ref{sec:cs} 
    & Uses four additional subcarriers from 802.11n; no influence on WiFi
    \gls{BER}; 4 QAM symbols per \gls{OFDM} symbol; max. covert rate 4.5 Mbit/s for 54\,Mbit/s WiFi frames.
    \\[.1cm]

    Cyclic Prefix\newline Replacement & Sec. \ref{sec:cp} 
    & (Partial) replacement of the cyclic prefix; no influence on WiFi
    \gls{BER} in line-of-sight channels, but affected by multiplath
    effects; 12 (full CP rep.)/6 (half CP rep.) QAM symbols per \gls{OFDM}
    symbol; max. covert rate 6.75 Mbit/s for $\nicefrac{1}{2}$ CP with CPCP
 \\[.1cm]
       	
	\hline
 
\end{tabular}
\caption{Summary of the analyzed covert channels. The exemplary performance values use our lab setup. Covert and\\ legitimate channel have a median raw \gls{BER} of below 0.1\% and use optimal settings for the covert channel.}   
\label{tab:sim_overview}
\end{table*}

\subsubsection{Detection}

\textbf{Layer 1.} A physical layer detector can compare the last 16 samples of
an \gls{OFDM} symbol with its cyclic prefix, which should be similar except for
\gls{ISI} damage.

Replacing parts of the original CP slightly increases
out-of-band emissions that might be visible in a spectrum analyzer---but they are still within the spectral mask
(see \autoref*{fig:cp-spectra}).

\textbf{Layer 2.} Since the \gls{CP} is removed before further processing on
Layer 2, the only visible effect is an increased \gls{BER} in rich multipath
environments. A Layer 2 detector cannot measure the actual
channel coefficients and thus, cannot distinguish whether a
high \gls{BER} is caused by a covert channel or not.

As expected, in the multipath channels the legitimate \gls{BER} significantly increases
at complete \gls{CP} replacement. However, we could not measure negative effects of $\nicefrac{1}{2}$ CP replacement
for channel models B, and D, as our results in \autoref*{fig:ber-wifi-sim-cpreplament} show.
In the practical measurements in \autoref*{fig:ber-wifi-mes-cpreplament}, only a full \gls{CP}
replacement has a negative effect on bit errors, especially when using off-the-shelf \glspl{NIC}.
Hence, attackers should replace less than $\nicefrac{1}{2}$ CP in typical environments
to avoid detection.

%% file: eval.tex
\section{Evaluation and Discussion}
\label{sec:evaluation}
Next, we compare results and discuss the pros and cons of the investigated covert channels, summarized in \autoref*{tab:sim_overview}.

All covert channels introduced in this paper can be combined. Since they modify different parts of \gls{OFDM} symbols, the overall performance when enabling all covert channels at once is their cumulative performance.
This comes at the cost of an increased detectability, see \autoref*{sec:detectability} on how to lower detectability.
If detected, Wendy either tries to decode the covert channel or to block it, for example, using a wireless firewall such as WiFire \cite{wifire}. 
The STF PSK covert channel is special, because even in case of detection it cannot be blocked.

\subsection{Covert Channel Performance}
A fair comparison of the covert channels is demanding, since they behave differently depending on the channel models, legitimate traffic, etc. 
The simulated \gls{AWGN} channel is overly optimistic compared to our lab setup, while channel model B is rather similar to our lab setup and yields comparable performance for the covert channels. Hence, we present empirical results for our lab measurements with a raw \gls{BER} of 0.1\%, which can easily be corrected with basic coding schemes.
Simulated channels D and E include effects not observable in our lab, hence yielding significantly harsher conditions for both covert and legitimate channel.

\begin{table}[!b]
\vspace{-1em}
\centering
\captionsetup{justification=centering}
\begin{tabular}{| l || c | c | c  | c |}
\hline
 							& STF 	& CFO 	& CS 	& CP \\
\hhline{|=||=|=|=|=|}
Layer 1 spectrum			&	n	&	n	&	y/n	&	y (p) \\
Layer 1 constellations		&	y (p)	&	y (p)	&	n 	&	y \\
Layer 1 decoding			&	n	&	n	&	y 	&	y \\
Layer 2 BER 				&	n 	&	y (p)	&	n 	&	y (p) \\
\hline
\end{tabular}
\caption{Detectability comparison: detectable (y), \\not detectable (n), detectability/performance trade-off (p).}
\label{tab:detectability}
\vspace{-1em}
\end{table}

Some covert channel rates are frame-based while others are symbol-based.
Depending on this, either the maximum  or minimum frame size is optimal to increase performance.
The minimum frame size is 14 bytes for \gls{CTS} and ACK frames.
Data frames can have a maximum frame size of up to 2338 bytes, assuming an unencrypted
802.11a/g data frame consisting of a \gls{MAC} header (typically 30 bytes),
a \gls{MSDU} (0-2304 bytes), and a \gls{FCS} (4 bytes) \cite{ieee-80211}.
Delays between frames depend on contention in the \gls{MAC} layer and on frame types,
hence we omit them in our exemplary calculation in \autoref*{tab:sim_overview}---as they are omitted when claiming an 802.11a/g maximum gross data rate of 54\,Mbit/s.
Choosing minimum or maximum frame size on Layer 2 might be suspicious to
attackers, thus this is only a reference for the optimal case. For low detection probability, the covert channel should be embedded in everyday network traffic.

\subsection{Detection Probability}
\label{sec:detectability}
\autoref*{tab:detectability} summarizes a comparison of the detectability of all the proposed covert channels.
Detectability is subject to the choice of the covert channel parameters; configuring the covert channel for lower throughput can facilitate to evade detection.

\textbf{Layer 1.}
A Layer 1 detector might take a look at the spectrum and IQ constellation diagrams with
a spectrum/signal analyzer. In case the Layer 1 detector must compare properties
in the time domain, a \gls{SDR} supported analysis is optimal.

In the spectrum, \gls{CP} replacement is visible since it introduces distortions
into the \gls{CP}, which violate a smooth signal continuation in it. 
camouflage subcarriers can be detected, but since their spectrum is valid for
802.11n, the signal field has to be decoded to identify the frame type.

When analyzing IQ constellations per symbol, all covert channels can be
detected. However, camouflage subcarriers can only be identified as such if the
signal field is decoded and checked. \gls{CP} replacement is visible in the
symbols after cutting off the \gls{CP}, when Wendy is in a multipath-rich
environment. Detection probability for STF PSK and CFO FSK can be lowered by
reducing $\Delta\varPhi$, respectively $\Delta$CFO.

\textbf{Layer 2}
A Layer 2 detector can only see an increasing \gls{BER}: if the covert channel is switched on and off immediately, \gls{BER} changes are visible on Layer 2.
Hence, STF PSK and camouflage subcarriers, which do not increase the normal channel \gls{BER},
are not detectable on Layer 2.
To reduce the detection probability of CFO FSK, reducing $\Delta$CFO helps. Replacing shorter parts of the \gls{CP} helps to diminish distortions in multipath-rich environments leading to lower overall \glspl{BER}.

%% file: related.tex
\section{Related Work}
\label{sec:relatedwork}

The idea of hiding information in wireless network traffic is not new. Most schemes are designed for the
data link layer or higher, using reserved fields, time delays, or packet corruptions.
An approach for transmitting data in corrupted frames was first proposed in~\cite{szczypiorski2003};
cryptographic information identifying corrupted frames is exchanged in advance using \gls{WEP} cipher \glspl{IV}
and MAC addresses.
\gls{WEP} \glspl{IV} are implemented in \cite{frikha2008}, but without making covert data match the same probability distribution as \glspl{IV}.
In~\cite{martins2010}, reserved fields are proposed for 802.15.4 covert channels.
An 802.11 MAC layer analysis on campus traffic in \cite{kraetzer2006} evaluated utilizable fields due to
randomness and high occurrence, proposing the \gls{FCF} More Frag, Retry, PwrMgt, More Data as well as the 802.11 header
fields Duration/ID and \gls{FCS}. In \cite{kraetzer2008}, timings of Retry bits indicating retransmissions are used to
encode information.
Hiding wireless access points by swapping fields with an Atheros and madwifi-ng is realized in~\cite{butti2006}.

Wireless physical layer covert channels are rare, but they are more generic.
Hence, related work in this area is not only on 802.11g but on \gls{OFDM} based systems in general.
In \cite{hijaz2010}, the usage of additional subcarriers in LTE and WiMAX is evaluated in simulation.
The model assumes that covert sender and normal sender are different identities, therefore their timing offset impacts
subcarrier orthogonality.
802.11n physical layer steganography using the \gls{CP} is proposed in \cite{grabski2013}.
In a simple \gls{AWGN} based simulation,
they archive a data rate as high as $\nicefrac{1}{4}$ of the normal channel without degradation.

To the best of our knowledge, there is only one wireless physical layer covert channel that
was put into practice:
dirty IQ constellations for 802.11a/g \cite{dutta2013}.
The authors define four IQ constellations in addition to the four raw QPSK points.
This way, they can reach up to the same covert throughput as normal throughput.
To circumvent detection,
they modified constellations to use a Gaussian distribution, and compared them to
regular noisy signals. However, when we tried to reproduce their results including the obfuscation mechanism, we had
to cope with a high amount of bit errors, especially in more complex channel models.

A related topic to covert channels is watermarking of signals, allowing for identification
and authorization on a physical layer basis. For this, an authentication tag is embedded. In \cite{tan2011cryptographic}, cognitive radio primary users add phase noise to QPSK symbols
to authenticate themselves while maintaining backward compatibility to secondary users who are not
aware of this scheme. A similar scheme for a non-return-to-zero encoding is proposed in \cite{kumar2014phy}
by embedding authentication tags in redundant information reducing \gls{ISI}.
A fingerprint can also be added to the channel state before sending, assuming only small channel changes
between transmissions, users knowing the previous channel state can extract the fingerprint \cite{goergen2010authenticating}.
The QPSK scheme is secured against user emulation attacks in \cite{borle2013physical} by adapting the phase
distortion to the current \gls{SNR}. However, all these schemes were only verified in simulations.
A practical implementation adding further IQ constellations as in \cite{dutta2013} without Gaussian distribution
is shown in \cite{yu2008physical}.

%% file: conclusion.tex
\section{Conclusion}
\label{sec:conclusion}

In this paper, we show that physical layer WiFi covert channels are feasible in practice. We design novel covert channels and improve known ones. Our work is---to the best of our knowledge---the first one to characterize various OFDM-based covert channels in practical settings. Based on our results, we discuss pros and cons of the covert channels with respect to their performance as well as their detectability. With this, we provide a first compendium for practical physical layer WiFi covert channels, which facilitates the understanding of this potential attack vector.